\documentclass[12pt]{article}
\usepackage{amsmath,amssymb,amsfonts}
\usepackage[pdftex]{graphicx}
\newcommand\beq{\begin{equation}}
\newcommand\eeq{\end{equation}}
\newcommand\bea{\begin{eqnarray}}
\newcommand\eea{\end{eqnarray}}

\title{\textbf{Line discontinuities, local action with both the field and its dual, and spin from
no spin in two-dimensional scalar theory}}

\author{Chandrasekhar Chatterjee$^{a}$\footnote{Current affiliation: CHEP, IISc, Bangalore 560012, India. 
E-mail: chandra@cts.iisc.ernet.in},
E. Harikumar$^{b}$\footnote{E-mail: harisp@uohyd.ernet.in},
Manu Mathur$^{c}$\footnote{E-mail: manu@bose.res.in},\\
Indrajit Mitra$^{d}$\footnote{E-mail: indrajit.mitra@saha.ac.in, imphys@caluniv.ac.in}~ and
H. S. Sharatchandra$^{a}$\footnote{E-mail: sharat@imsc.res.in} \\\\
$^a$ The Institute of Mathematical Sciences, C.I.T. Campus, Taramani P.O.,\\
Chennai 600113, India\\
$^b$ School of Physics, University of Hyderabad, Central University P.O.,\\
Hyderabad 500046, India\\
$^c$ S. N. Bose National Centre for Basic Sciences,\\
JD Block, Sector III, Salt Lake City, Calcutta 700098, India\\
$^d$ Department of Physics, University of Calcutta,\\
92 A.P.C. Road, Kolkata 700009, India}
\date{}

\begin{document}

\maketitle


\begin{abstract}

We consider a local action with both the real scalar field and its dual in two Euclidean dimensions.
The role of singular line discontinuities is emphasized. Exotic properties of the correlation
of the field with its dual, the generation of spin from scalar fields, and quantization of dual
charges are pointed out.  Wick's theorem and rotation properties of fermions are recovered for
half-integer quantization.  

\end{abstract}

\section{\bf Introduction}
\label{sec1}

Duality transformations play important roles in many field theories and statistical physics 
models. They provide an equivalent description of the system.  They often relate two models 
which appear to be different. They map weak coupling or high temperature regime of one model to the 
strong coupling or low temperature regime of the other. This provides an understanding of the phases 
of the model  as  well as  non-perturbative calculational techniques. The degrees of freedom of the dual theory 
play the role of disorder variables of the original theory.  They often provide valuable order parameters 
to characterize the phases of the model.  In addition to all this, order and disorder variables close 
together have exotic properties. For instance, certain combinations of bosonic variables may behave as 
a fermion.  Such combinations also play a crucial role in providing the description and properties of 
the model in some phases. Because of these important features, it is useful to have  systematic 
techniques for handling both the field and its dual on an equal footing. In this work we do this for 
a free massless real scalar field in two (Euclidean) dimensions. 
This case, though simple, 
encodes rich physics. It is self-dual, so that the dual field is also free scalar field. 
We also want to handle correlation
function of the field with its dual.
We thus obtain a {\it local} action which has both
the fields simultaneously present. We derive
this action in the presence of source terms
which bring out the role of line
discontinuities in the field configurations.
We emphasize how the line discontinuities
result in novel properties of the mixed correlation functions,
and explain how they lead to the generation of spin from scalar fields,
the quantization of dual charges and the construction of the composite
fermion operator. The local action involving both the field and its dual was
studied in the context of string theory in Refs.\ \cite{tseytlin1,tseytlin2}.
The action is also
related to Hamiltonians used for  describing condensed matter systems in one dimension
\cite{kane1,kane2}.

In Section \ref{sec2.1}, we express the partition function in terms of the dual field.
In Section \ref{sec2.2}, we bring out the role of line discontinuities, in close analogy with the order-disorder 
formulation of the 2-d Ising model \cite{kadceva}. We emphasize that such discontinuities in a field configuration are 
meaningful without requiring a smoothening, and contribute a finite action. In Section \ref{sec2.3}, we
arrive at  the local action with both the field and its dual present.
This local action is not 
manifestly invariant under rotations. The underlying reason is the choice of the line of discontinuity. 
It is also not real in spite of being equivalent to the free massless scalar theory. Even though it has double 
the number of fields, it represents the original theory. 

The correlation of order and disorder variables has been discussed in various 
contexts in \cite{marino1,marino2}. In our case, the correlation function of the field 
with its dual has a number of exotic properties, which we  explain in
Section \ref{sec2.4}. First, it is discontinuous, 
reflecting that the field introduces a line of discontinuity in the configuration of dual field. The 
correlation function does not die off at large distances, representing the non-local effect the field has. 
Indeed the correlation function  is proportional to the angle subtended by the vector joining the coordinates 
of the field and the dual field with the line of discontinuity. Both the field and its dual being scalars,
the correlation functions involving only the field (or only the dual field) 
depend on the distance and not on the orientation. This is no longer true of the mutual correlations, again 
 reflecting  the role of line discontinuities. In addition to all this, the correlation function is pure 
imaginary even though the fields are real. This is traced to the action which is not real.

The free massless real scalar field in two Euclidean dimensions embodies all minimal models of conformal 
field theory through the use  of boundary charges, the Coulomb gas formalism and the vertex 
operators \cite{ketov,cft}. In Section \ref{vopdp}, we show that  
the vertex operators of both the field and its dual can be handled together
in the formalism we use. This exhibits the phenomenon 
of `spin from no spin'.  The correlation functions imply that the vertex operators of the dual field 
(or equivalently the original field) acquire spin though constructed from a scalar field.  The mixed 
correlation functions appear to be multivalued.  But we emphasize that they are simply discontinuous, a 
consequence of line discontinuities. For a specific `quantization' of the `charges' of the field and its
dual, the single valuedness of the correlation function is restored.  This is exactly parallel to Dirac 
quantization of electric and magnetic charges. Just as the Dirac string becomes invisible, the line discontinuities 
become invisible for the specific vertex operators which acquire spins.  The correlation functions are now 
single valued and rotation covariant with spins.

The composite operators built out of order and disorder fields often have exotic properties and play 
important roles in the properties of the theory. In 2-d Ising model, the combination is a fermion 
as regards  to rotational properties and spin-statistics connection 
\cite{kadceva,d=2(1),d=2(2)}.
Analogous to this is the 2-d 
bosonization, which builds a fermion from real massless boson 
\cite{mandel,boson1,boson2,boson3,boson4,boson5}. The  formalism used in our work is well equipped 
to handle this. In Section \ref{ffb}, we consider half-integer quantization 
of the dual charges and show how Wick's theorem 
for fermions is recovered.  Here again we point out the role played by the line discontinuities to yield the rotational 
properties of the fermions. 
The prescription for handling Mandelstam's \cite{mandel} composite operator for the fermion is given.
We also offer new insight into the Klein factors 
\cite{klein} required in bosonization.

We discuss our results in Section \ref{Discussion}. Here we indicate how the line discontinuity is related to
the soliton of the sine-Gordon theory. We also point out how to write the
local action with both the field and its dual in the presence of self-interactions.

We choose straight line  discontinuities through most of the paper. However, we have the freedom 
of choosing any path, as discussed  in the Appendix.

\section{A local action with both the field and its dual, and the role of
line discontinuities}\label{sec2} 

\subsection{Partition function in terms of dual field}\label{sec2.1}

Consider a free massless scalar field $\phi(x)$ in two Euclidean dimensions. Its correlation functions 
can be obtained from the generating functional: 
\bea 
Z[\rho] = {\cal N}_1 \int {\cal D} \phi e^{\int d^2x\left[-\frac{1}{2} \sum_{i=1}^{2} 
\left(\partial_i\phi(x)\right)^2 + i \rho(x) \phi(x)\right]} 
\label{pf} 
\eea 

In (\ref{pf}) $\rho(x)$ is the source coupling locally to $\phi(x)$ and 
the normalization ${\cal N}_1$ is defined so that $Z[\rho=0] =1$.  We are inserting 
$\sqrt{-1}$ in the source term for two reasons: (i) the duality transformation carried out below 
is more transparent, (ii) the point sources\footnote{Throughout this paper, we use
superscripts to label different  points in 2-d space.}
\bea 
\rho(x) = \sum_i e_i \delta^2(\vec x-\vec{x}^{i})  
\label{pos} 
\eea 
give the vertex operators 
\bea  
V_i = e^{ie_i\phi(x^i)} 
\label{vo} 
\eea 
which play crucial role here as in CFTs. Formally: 
\bea 
Z[\rho] = e^{-\frac{1}{2} \int d^2x d^2y \rho(x) \bigtriangleup (x-y) \rho(y)}, 
\label{pf2} 
\eea 
where  $\bigtriangleup (x-y)$ is the 2-d Coulomb potential satisfying: 
\bea 
- \partial^2_{x} \bigtriangleup (x-y) =  \delta^2(\vec x - \vec y). 
\label{cp} 
\eea  
As the Coulomb potential in 2-d is infrared divergent, we use a finite area of linear dimension 
$R$. Also to handle ultraviolet divergences in certain correlation functions, we use the ultraviolet cutoff $a$ 
to get \cite{tsvelik}: 
\bea 
\bigtriangleup (x) = \frac{1}{4\pi} 
\ln \left(\frac{R^2}{\vec{x}^2 + a^2}\right). 
\label{ps} 
\eea 
Using an auxiliary field $\vec{b}_i(x)$ we linearize the $\phi(x)$ terms in (\ref{pf}): 
\bea
Z[\rho] = {\cal N}_2 \int {\cal D} \phi {\cal D} b_i e^{\int d^2 x 
\left(-\frac{1}{2} b_i^2(x) + i b_i(x) \partial_i\phi(x) + i \rho(x) \phi(x)\right)}. 
\label{lpf1} 
\eea 
A formal integration over $\phi$ gives a functional $\delta$ function: 
\bea 
Z[\rho] = {\cal N}_3 \int {\cal D} b_i \prod_x\delta\left(\partial_i b_i(x) - \rho(x)\right) 
e^{-\frac{1}{2} \int d^2 x b_i^2(x)}.  
\label{dpf} 
\eea 
We solve the `Gauss law constraint' as follows: 
\bea 
b_i(x) = \epsilon_{ij} \partial_j \tilde \phi(x) + \delta_{i1} \partial_1^{-1} \rho(x)
\label{sgl} 
\eea 
where $\partial_1^{-1}$ is defined as: 
\bea 
\partial_1^{-1} \rho(x) = \int_{- \infty}^{x_1} d x_1' \rho(x_1',x_2). 
\label{di} 
\eea 
For a point source $\rho(x) = \delta^2(x-x^0)$, 
\bea 
\partial_1^{-1} \rho(x) = \theta(x_1-x_1^0) \delta(x_2-x_2^0), 
\label{lds}
\eea
i.e. the flux is carried along a line from $x^0$ parallel to $x_1$ axis in the positive direction. 
The `Gauss law' solution (\ref{sgl}) implies, 
\bea 
Z[\rho] = {\cal N}_4 \int {\cal D} \tilde \phi e^{-\frac{1}{2} \int d^2x \left(\left(\partial_1 \tilde \phi(x)\right)^2
+ \left(\partial_2 \tilde \phi(x) + \partial_1^{-1} \rho(x)\right)^2\right)}.  
\label{pfwf}
\eea 
This re-expresses the partition function (\ref{pf}) in terms of a new massless scalar field $\tilde \phi(x)$ which 
is dual of the field $\phi(x)$.  The source $\rho$, which couples locally to $\phi$, 
couples in a specific non-local manner to the dual field $\tilde \phi$.  
In spite of this, an integration over 
$\tilde \phi$ 
reproduces (\ref{pf2}): 
\bea 
Z[\rho] & = & {\cal N}_4 \int {\cal D} \tilde \phi e^{\int d^2 x \left[-\frac{1}{2} \left(\partial_i
\tilde \phi(x)\right)^2 + \tilde \phi(x) \partial_2 \partial_1^{-1} \rho(x) 
-\frac{1}{2} \left(\partial_1^{-1}\rho(x)\right)^2\right]} \nonumber \\
&=& e^{ \frac{1}{2} \left[\int d^2x d^2y \partial_2\partial_1^{-1} \rho(x) \bigtriangleup(x-y)  
\partial_2\partial_1^{-1} \rho(y)  - \int d^2 x \left(\partial_1^{-1}\rho(x)\right)^2\right]}. 
\eea 
Now integrating by parts with respect to $x_2$ and $y_2$, 
and using 
$ \left(\frac{\partial}{\partial x_2}\right)^2  \bigtriangleup(x-y) = 
-\left(\frac{\partial}{\partial x_1}\right)^2 \bigtriangleup(x-y) - \delta^2(x-y)$ and $\partial_1\partial_1^{-1} 
\rho(x)=\rho(x)$, we recover (\ref{pf2}). 

\subsection{Interpretation in terms of  lines of discontinuity} 
\label{sec2.2}

For the point sources (\ref{pos}) the partition function (\ref{pfwf}) takes the form: 
\bea 
Z(e_i) = {\cal N}_4 \int{\cal D}\tilde \phi e^{-\frac{1}{2} \int d^2x \left[\left(\partial_1\tilde \phi(x)\right)^2 + 
\left(\partial_2 \tilde \phi(x) + \sum_i e_i \theta (x_1-x^i_1) \delta(x_2-x_2^i)\right)^2 \right]}. 
\label{pspf} 
\eea 
This exhibits the non-local interaction of the dual field $\tilde \phi(x)$ with the local sources of $\phi(x)$. 
A point source at $\vec x^i$ of strength $e_i$ corresponds to a singular flux line 
which starts at $\vec x^i$ and goes parallel to $x_1$ axis in the 
positive direction.  Despite the singular line, the partition function (\ref{pspf}) is meaningful 
and reproduces (\ref{pf2}) as seen above. The singular line forces the configurations $\tilde \phi(x)$ 
which contribute to the partition function (\ref{pspf}) to have discontinuities 
along the singular line, so as to cancel the singularities in the exponent and contribute a finite action. 
Thus {\it the configurations $\tilde\phi(x)$ which contribute to the partition function are no longer smooth and continuous 
configurations, but those with specific lines of discontinuity.}

\subsection{A local action with both the field and its dual present}\label{sec2.3}

Using an auxiliary field $\tilde \pi(x)$, the partition function  (\ref{pfwf}) can be written as: 
\bea 
Z[\rho] = {\cal N}_5 \int {\cal D} \tilde \phi {\cal D} \tilde \pi e^{\int d^2x \left[-\frac{1}{2}
\left(\partial_1\tilde \phi(x)\right)^2 -\frac{1}{2} \tilde \pi^2(x) + i \tilde \pi(x) \left(\partial_2\tilde \phi(x) 
+ \partial_1^{-1}\rho(x)\right)\right]}.  
\label{pf3} 
\eea 

Notice that $\rho$ couples locally to $- \partial_1^{-1} \tilde \pi$. Therefore, we identify 
\bea 
- \partial_1^{-1} \tilde \pi(x) \equiv \phi(x). 
\label{fp} 
\eea 
We can now explore the correlation function of $\phi(x)$ with its dual field $\tilde \phi(x)$. 
For this we introduce a local source $\tilde \rho(x)$ and couple it to $\tilde \phi(x)$. The partition function 
(\ref{pf3}) takes the form: 
\bea 
Z[\rho,\tilde \rho] = {\cal N}_6 \int {\cal D} \phi {\cal D} \tilde \phi 
e^{\int d^2x \left[-\frac{1}{2} 
\left(\partial_1\phi(x)\right)^2- 
\frac{1}{2} \left(\partial_1\tilde \phi(x)\right)^2 -i \partial_1\phi(x) \partial_2\tilde \phi(x) 
+ i \left(\rho(x) \phi(x) + \tilde \rho(x) \tilde \phi(x)\right)\right]}. 
\label{pfrc}
\eea 
The partition function (\ref{pfrc}) represents the original theory (\ref{pf}) with only $\phi(x)$ field 
as an equivalent local theory with both the field and  its dual simultaneously present.
Moreover it is self-dual as the action itself is invariant
under:
$\phi \leftrightarrow \tilde \phi$ and $\rho \leftrightarrow \tilde \rho$.
The action in (\ref{pfrc}) also has discrete inversion symmetry:
$\phi \rightarrow-\phi,~ \tilde \phi \rightarrow -\tilde \phi,~ \rho \rightarrow - \rho,~ \tilde \rho \rightarrow -\tilde \rho$.
In the absence of sources, the action  in (\ref{pfrc}) is also
invariant under the following global transformations:
\bea
\phi(x_1,x_2) \rightarrow \phi(x_1,x_2) + \sigma, ~~~ \tilde \phi(x_1,x_2) \rightarrow \tilde \phi(x_1,x_2)
+ \tilde \sigma.
\label{sgt}
\eea
However, it is not manifestly 
invariant under rotations of the 2-d space. The traditional $(\partial_2\phi)^2$ and $(\partial_2\tilde \phi)^2$ 
terms are not present. Also, the action is not real (even in the absence of the imaginary sources we are 
using). The  term bilinear in $\phi$ and $\tilde \phi$ has $\sqrt{-1}$ factor.
(This $\sqrt{-1}$ factor, however, would not be present in Minkowski space. This is the case considered in Refs.\
\cite{tseytlin1,tseytlin2}.)
  In spite of all this, the equivalence of the action to  
the traditional partition function is seen by integrating over one or the other field (in the absence of sources).
It should be noted that  
{\it the lack of manifest rotation invariance can be traced back to the specific choice of the $x_1$-direction 
in our solution (\ref{sgl}) of the Gauss law constraint.} 

We see from (\ref{pf3}) that if we regard $x_2$ as the Euclidean time, the partition function 
(\ref{pfwf}) corresponds to the Hamiltonian density $(Z=\lim_{T \rightarrow \infty} {\rm Tr} e^{-TH})$: 
\bea 
{\cal H}(\tilde \pi,\tilde \phi) = \frac{1}{2} \tilde \pi^2(x) + \frac{1}{2} \left(\partial_1\tilde \phi(x)\right)^2 
\label{hd}
\eea 
with  canonically conjugate variables $(\tilde \phi(x),\tilde \pi(x))$. Thus (\ref{fp}) relates the field $\phi(x)$ 
non-locally to the momentum $\tilde \pi$ conjugate to the dual field $\tilde \phi(x)$ \cite{senechal}. This connects 
the formalism of the present work to condensed matter system Hamiltonians \cite{kane1,kane2}. 
With this interpretation, (\ref{pfrc}) is not so exotic. Nevertheless, it is very useful as the 
field and its dual are on the same footing.

\subsection{Correlations of the field $\phi$ with its dual $\tilde \phi$}\label{sec2.4}

Evaluating (\ref{pfrc}) and taking functional derivatives with respect to the sources $\rho(x)$ and $\tilde \rho(x)$ 
we get correlation functions of field $\phi(x)$  with its dual $\tilde \phi(x)$: 
\bea 
Z[\rho,\tilde \rho] & = &  {\cal N}_6 \int {\cal D}\phi {\cal D}\tilde \phi e^{\int d^2x 
\Big[-\frac{1}{2} \left(\phi(x)~ \tilde \phi(x)\right) 
{\tiny {\left(\begin{array}{cc} -\partial_1^2  & -i \partial_1 \partial_2 \\ -i\partial_1\partial_2  
& -\partial_1^2  \end{array}\right) ~\left(\begin{array}{c}\phi(x)\\ \tilde \phi(x) \end{array}\right)}} 
+ i \left(\phi(x)~ \tilde \phi(x)\right){\tiny{\left(\begin{array}{c}\rho(x)\\ \tilde \rho(x) \end{array}\right)}}\Big]}, 
\nonumber \\ 
&=& e^{- \frac{1}{2} \int d^2x d^2y \left[ \left(\rho(x)~ \tilde \rho(x)\right) 
{\tiny{\left(\begin{array}{cc} - \partial_1^2  & -i \partial_1 \partial_2 \\ - i\partial_1\partial_2  
& - \partial_1^2  \end{array}\right)}} ^{-1} 
\hspace{-.3cm}(x,y) ~
{\tiny \left(\begin{array}{c}\rho(y)\\ \tilde \rho(y) \end{array}\right)}\right]}. 
\label{zrj} 
\eea     
Now using  
$$\Big(\begin{array}{cc}  - \partial_1^2  & - i \partial_1 \partial_2 \\ -i\partial_1\partial_2  
& -\partial_1^2  \end{array}\Big)^{{}^{-1}}\hspace{-.3cm}(x,y) = 
\Big(\begin{array}{cc}  1   & -i \partial_1^{-1} \partial_2 \\ -i\partial_1^{-1} \partial_2  
&  1  \end{array}\Big) \bigtriangleup(x-y),$$ 
we get: 
\bea 
\langle\phi(x) \phi(y) \rangle = \bigtriangleup(x-y) 
\label{ppp} 
\eea 
which is consistent with (\ref{pf2}), and 
\bea 
\langle \tilde \phi(x) \tilde \phi(y) \rangle = \bigtriangleup(x-y) 
\label{ccp} 
\eea 
as expected for a massless real scalar field.  The correlation function 
of the field $\phi(x)$ with its dual $\tilde \phi(y)$ is: 
\bea 
\langle \phi(x) \tilde \phi(y) \rangle = 
-\frac{i}{2} [\partial_1^{-1} \partial_2 \bigtriangleup(\vec x - \vec y) + ~x \leftrightarrow y~].  
\label{pcp} 
\eea 
In (\ref{pcp}) the operator $\partial_1^{-1}\partial_2$ acting on $\bigtriangleup(\vec x - \vec y)$ 
is with respect to the first coordinate $\vec x$.
We define:  
\bea 
\frac{{\Theta}(\vec x)}{2\pi} \equiv 
 \partial_1^{-1} \partial_2 \bigtriangleup(x) =  - \frac{1}{2\pi} \partial_1^{-1} \frac{x_2}{r^2} = 
 -\frac{x_2}{2\pi} \int_{-\infty}^{x_1} dx_1' \frac{1}{(x_1')^2+(x_2)^2} = -\frac{1}{2\pi} \left(\tan^{-1}\frac{x_1}{x_2} 
		 +\frac{\pi}{2}\right).    
\label{tt} 
\eea 
We now have to directly 
address the integral to choose the branch for the definition  of inverse of tan. The integral 
\bea 
\int_{-\infty}^{x_1} dx_1' \frac{1}{(x_1')^2 + (x_2)^2}  
\label{dsp} 
\eea 
is positive definite and well defined as long as $x_2 \neq 0$. Even for $x_2=0$, it is well defined if 
$x_1 < 0$ because the integrand never blows up. Making a change of variables from $x_1'$ to $ t \equiv 
\frac{x_1'}{|x_2|}$, we get 
\bea 
\Theta(\vec x) = -{\rm sgn}(x_2) \int_{- \infty}^{\frac{x_1}{|x_2|}} dt \frac{1}{1+t^2} 
\label{pap} 
\eea
where ${\rm sgn}(x_2)$ equals $+1$($-1$) for $x_2 >0$ ($x_2<0$). 
\begin{figure}[t]
\centering
\includegraphics[width=14cm,height=7cm]{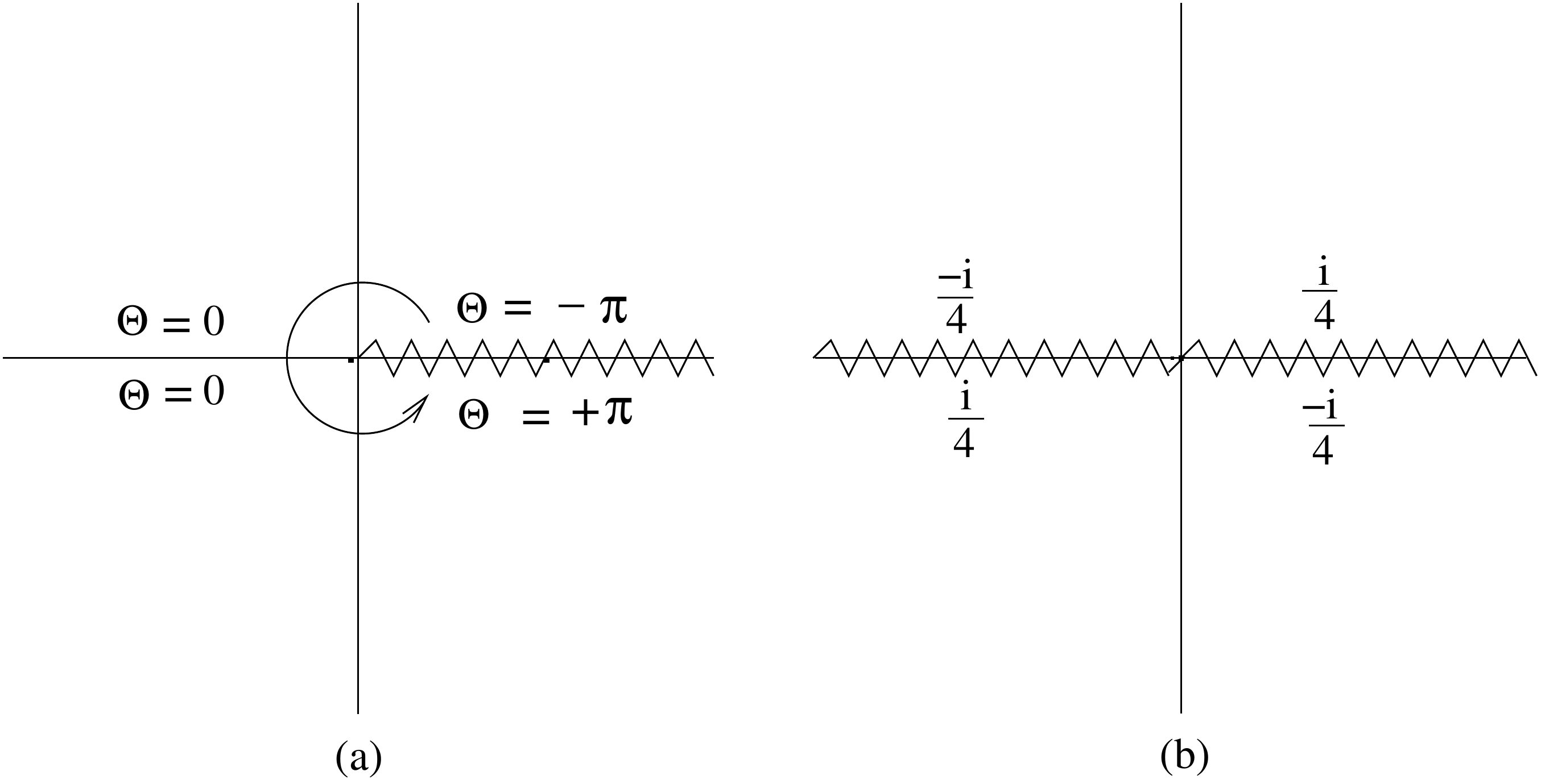}
\caption{The discontinuities in (a) the angle $\Theta(\vec x)$, (b) the correlation function 
$\langle \phi(x) \tilde \phi(0)\rangle = - \frac{i}{4\pi} \left(\Theta(\vec x ) + 
\Theta (-\vec x)\right)$ across the (horizontal) $x_1$-axis.} 
\label{fig1}
\end{figure}
Therefore, as shown in Figure \ref{fig1},  $\Theta(\vec x)$ is the angle subtended by the 
vector $\vec x$ measured in the 
anti-clockwise direction with starting value $-\pi$ from the positive $x_1$-axis.  Thus 
\bea 
\langle\phi(x)\tilde \phi(y)\rangle =  
- \frac{i}{4\pi} \left(\Theta(\vec x -\vec y) + \Theta (\vec y -\vec x)\right).
\label{ang} 
\eea 
Note that $\Theta(\vec x)$  is discontinuous along the positive $x_1$ axis because it changes from $-\pi$ 
to $+\pi$ as $x_2$ changes from a positive value to negative value as shown in Figure \ref{fig1}-a. Thus the correlation 
function of the field with its dual (\ref{ang}) is discontinuous whenever $\vec x - \vec y$ is parallel to the 
$x_1$-axis.  As shown in Figure \ref{fig1}-b, it changes by 
$\frac{i}{2}{\rm sgn}(x_1-y_1)$ as $(x_2-y_2)$ changes sign from negative to positive.
Note that the $x_1$-axis is singled out in (\ref{sgl}), leading to $\partial_1^2$  in the diagonal elements in (\ref{zrj}),
and then to $\partial_1^{-1}$ in (\ref{pcp}). Thus {\it this discontinuity is a reflection of the 
boundary condition forced by the line discontinuity 
generated by the field $\phi(x)$ in the configuration space 
of its dual field $\tilde \phi(y)$.} 

It is convenient to define the standard azimuthal angle $\theta(\vec x)$ 
$(0 < \theta(\vec x) < 2\pi)$ by 
\bea 
\theta(\vec x) \equiv \Theta(\vec x) + \pi,
\label{nth}
\eea  
obeying the standard relation: 
\bea
\theta(-\vec x) = \theta(\vec x) + \pi {\rm sgn}(x_2).
\eea 
The correlation function (\ref{ang}) can now be written as: 
\bea 
\langle\phi(x)\tilde \phi(y)\rangle =  
- i \left(\frac{\theta(\vec x -\vec y)}{2\pi}  + \frac{{\rm sgn}(x_2-y_2)}{4} -\frac{1}{2}\right).
\label{angn} 
\eea 
Thus the correlation function of the field $\phi(x)$ with its dual $\tilde \phi(y)$ has the following unusual 
properties: 
\begin{enumerate} 
\item It does not fall off with the distance,  reflecting the non-local effect $\phi(x)$ has 
in the configuration space of $\tilde \phi(x)$. 
\item {\it The correlation depends on the orientation of the relative vector $\vec x - \vec y$  in space.}
Note that the fields $\phi(x)$ and $\tilde \phi(x)$ are by themselves scalar fields and therefore their 
correlations would be expected to  depend only on the distance of separations and not on the orientation 
of the relative vector. This scalar nature is true if we consider the correlations  of the field $\phi$ only
(or $\tilde \phi$ only) at various points. However,  as discussed in detail in Section \ref{vopdp},  
the correlations of 
$\tilde \phi$ and $\phi$  acquire  a  direction dependence.   
{\it The reason again is the non-local effect produced by $\phi$ in the 
configurations of $\tilde \phi$ or vice-versa.} 
\item Even though $\phi$ and $\tilde \phi$ are real fields, their mutual correlations are pure imaginary. The technical 
reason is that the action in the partition function (\ref{pfrc}) is no longer real. 
\item The $\langle \phi(x) \tilde \phi(y)\rangle$  correlations given by (\ref{ang}) are symmetric under the interchange 
of $\vec x$ and $\vec y$. The reason can be traced to  (\ref{zrj}) where we inverted a symmetric matrix.
\end{enumerate} 
We finally have the master formula (see equations (\ref{zrj})-(\ref{pcp})): 
\bea
Z[\rho,\tilde \rho]=e^{\int d^2x d^2y \left[-\frac{1}{2}\rho(x) \bigtriangleup(x-y) \rho(y)  
-\frac{1}{2}\tilde \rho(x) \bigtriangleup(x-y) \tilde \rho(y) + {i} \rho(x) 
\left(\frac{\theta(\vec x -\vec y)}{2\pi}  + \frac{{\rm sgn}(x_2-y_2)}{4} -\frac{1}{2}\right)
\tilde \rho(y)\right]}.
\label{maeq} 
\eea

One can also use the linear combinations 
of the fields $\phi(x)$ and its dual $\tilde \phi(x)$: 
\bea 
\xi_{\pm}(x) = \frac{1}{2}\left(\phi(x) \pm \tilde \phi(x)\right). 
\eea 
The action in (\ref{pfrc}) decouples in $\xi_+$ and $\xi_-$  \cite{tseytlin1,tseytlin2}: 
\bea 
S[\phi,\tilde \phi] = S\left[\xi_+,\xi_- \right] = 
\int d^2x \left(\left(\partial_1\xi_+\right)^2 + i \partial_1\xi_+\partial_2\xi_+ 
+\left(\partial_1\xi_-\right)^2 - i \partial_1\xi_-\partial_2\xi_-\right)\,.
\eea 
The correlation functions of these combinations are: 
\bea 
\langle \xi_+(x) \xi_+(0)\rangle &= &-\frac{1}{4\pi} \left(\ln \frac{z}{R} 
+ i\pi \left(\frac{{\rm sgn}~ x_2}{2}-1\right)\right), \nonumber \\
\langle \xi_-(x) \xi_-(0)\rangle &= &-\frac{1}{4\pi} \left(\ln \frac{\bar z}{R} 
- i\pi \left(\frac{{\rm sgn}~ x_2}{2}-1\right)\right),
\label{ccf}
\eea 
where $z=x_1+ix_2$.  The fields $\xi_{\pm}$ are related to holomorphic and anti-holomorphic 
fields $\varphi(z)$ and $\bar\varphi(\bar z)$ used in CFT, and are involved in bosonization also. 
(See, for example, Refs.\ \cite{tsvelik} and \cite{senechal}; see also our equation (\ref{identn}).) 

\section{ Vertex operators of the field $\phi$ and its dual $\tilde \phi$}
\label{vopdp} 

Point sources (\ref{pos})  correspond to vertex operators $e^{ie_i \phi(x^i)}$ of well known Coulomb gas 
formalism for minimal CFTs \cite{ketov,cft}. Their well known correlation functions are read off from the  master 
formula (\ref{pf2}) and the regularized propagator (\ref{ps}),

\begin{eqnarray}
\langle\prod_i e^{ie_i\phi(x^i)}\rangle&=&e^{-\frac{1}{2}\int d^2x\,d^2y(\sum_i e_i\delta^2(x-x^i)) 
\Delta(x-y)(\sum_j e_j\delta^2(y-x^j))}\nonumber\\
&=&\left(\frac{R}{a}\right)^{-\frac{1}{4\pi}(\sum_i e_i)^2}
\prod_{i<j}\left(\frac{|x^i-x^j|}{a}\right)^{\frac{e_i e_j}{2\pi}}.   
\label{R/a}
\end{eqnarray}
When $R\to\infty$, only the net charge zero ($\sum_i e_i=0$) correlation functions
survive. Then, using 
$\sum_{i<j}e_i e_j=-\frac{1}{2}\sum_i e_i^2$,
we can write the $a$-dependence in Eq.\ (\ref{R/a}) as
$\prod_ia^{e_i^2/4\pi}$.
So, to get finite correlations as the ultraviolet regulator $a\to 0$, we `renormalize' the vertex 
operators \cite{tsvelik}:
\begin{equation}
: e^{ie_i\phi(x)}:= a^{-\frac{e_i^2}{4\pi}} e^{ie_i\phi(x)}.
\label{ren} 
\end{equation}
This amounts to removing self correlations $\langle\phi(x)\phi(x)\rangle$ in $e^{ie\phi(x)}$. 
The expression in Eq.\ (\ref{R/a}) then reduces to just
$\prod_{i<j}|x^i-x^j|^{\frac{e_i e_j}{2\pi}}$.
For the neutral combination,
\begin{equation}
\langle:e^{ie\phi(x)}: :e^{-ie\phi(y)}:\rangle=\frac{1}{|x-y|^{\frac{e^2}{2\pi}}}.
\label{ppco} 
\end{equation}
We may define vertex operators for the dual field $\tilde \phi$ in a similar manner:
\begin{equation}
\langle :e^{ig\tilde \phi(x)}: : e^{-ig\tilde \phi(y)}:\rangle=\frac{1}{|x-y|^{\frac{g^2}{2\pi}}}.
\label{ccco} 
\end{equation}

%
\noindent The mixed correlation functions can be computed using the master formula (\ref{maeq}). We need 
charge neutrality separately for field $\phi$ and dual field $\tilde \phi$, for finite correlation function 
in $R\to\infty$ limit. It is as if we have two Coulomb gas partition functions, except that 
cross-correlations are also present:
\begin{equation}
\langle\prod_i :e^{ie_i\phi(x^i)}: \prod_{k}: e^{ig_k \tilde \phi(y^k)}: \rangle =
\prod_{i<j} |{\vec x}^i-{\vec x}^j|^{\frac{e_ie_j}{2\pi}}
\prod_{k<l} |\vec y^k-\vec y^l|^{\frac{g_kg_l}{2\pi}} 
\prod_{i,k} e^{ie_i g_k
\left[\frac{\theta(\vec x^i -\vec y^k)}{2\pi}  + \frac{{\rm sgn}(x^i_2-y^k_2)}{4} -\frac{1}{2}\right]}
\label{pcrc} 
\end{equation}
with $\sum_i e_i=0,\sum_k g_k=0$.

Generally  these correlation function have jumps in phase  whenever a relative vector 
${\vec x}^i-{\vec y}^k$ tends to be parallel to the $x_1$-axis (see Figure \ref{fig1}-b). 
We may alternatively interpret 
that the correlation functions are multivalued.
The discontinuity, as explained after (\ref{ang}), reflects the line discontinuity 
produced by $\phi$ in the 
configuration space of $\tilde \phi$. The specific role of $x_1$-axis is due to our choice of the 
solution of the Gauss' law constraint. {\it This parallels the role of Dirac string in Dirac formulation 
of the magnetic monopole.} Let $e_i=m_i e, g_k=n_k g$ (where $m_i$ and $n_k$ are integers) with the 
Schwinger quantization condition 
\begin{equation}
eg=4\pi\,.
\label{dqc} 
\end{equation}
Consider the lowest values $m_i=1$ and $n_k=1$. The corresponding phase
in (\ref{pcrc}) is then $\exp[-\,4\pi\langle\phi(x)\tilde\phi(y)\rangle]$,
and equals $e^{\pm i\pi}$ across the $x_1$-axis
(refer to Figure \ref{fig1}-b).
Thus the phase is continuous, the `string' is no longer visible and 
the correlation functions are single valued.

We now briefly explain when we get the Schwinger quantization condition, and when the Dirac 
quantization condition $eg=2\pi$.
By introducing field $\phi$ in (\ref{pfrc}) and inverting the symmetric matrix in (\ref{zrj}), we got 
$\frac{1}{2}( \Theta(\vec x-\vec y)+\Theta(\vec y-\vec x))$ in (\ref{ang}).
This is the analogue of the 
Schwinger potential, since  the line discontinuity is
distributed to both positive  and negative $x_1$  axis ( as in Figure \ref{fig1}-b).
Thus this is due to insisting on a symmetric quadratic form in $\phi$ and $\tilde\phi$.
On the other hand, having only $\Theta(\vec x-\vec y)$ in (\ref{ang}) is the analogue of the Dirac potential, since it
has the line discontinuity only across the positive $x_1$  axis. This
 is obtained from the partition function in terms of the dual field $\tilde\phi$ alone,
after adding a source
for $\tilde\phi$  and integrating over $\tilde\phi$. This has been done in the Appendix.
In this case the `string' becomes invisible for
the Dirac quantization condition $eg=2\pi$.
 
Rotation covariance now manifests in an unusual manner. Though $\phi$ and $\tilde \phi$ 
were separately scalar fields, the mutual correlation functions (\ref{pcrc}) undergo transformation 
under the rotation $\theta({\vec x-\vec y})\to \theta({\vec x-\vec y})+\omega$. 
In other words, (\ref{pcrc}) are not invariant, but only covariant, under rotation. 
Thus the  vertex operators 
have acquired a `spin'. This is in analogy with the Saha-Wilson \cite{sahawilson1,sahawilson2,sahawilson3} contribution (of 
the electromagnetic field) to the total angular momentum of a charge-monopole system.
With $eg=2\pi$, (\ref{pcrc}) transforms under the above rotation by a factor of $e^{\pm i\omega}$. So
the vertex operators may be assigned the following transformation property under rotation:
\begin{eqnarray}
e^{\pm ie\phi(x)}&\to& e^{\pm ie\phi(x)},\nonumber\\
e^{\pm ig\tilde \phi(y)}&\to& e^{\pm i\omega}e^{\pm ig\tilde \phi(y)},
\label{tra} 
\end{eqnarray} 
i.e., the dual field vertex operators $e^{\pm ig\tilde \phi(y)}$ have acquired spin $\pm$ 1. 
Equivalently we may assign spin to $e^{\pm ie\phi(x)}$.
We emphasize that {\it the underlying reason for the generation of spin from scalar fields is that
the correlation function between $\phi(x)$ and $\phi\tilde(y)$  depends on the orientation of the
relative vector $\vec x-\vec y$.}

\section{Fermions from Bosons} 
\label{ffb} 

In the previous section, we obtained vertex operators with spin from scalar fields. With 
$\frac{e g}{2\pi} = 1$, the spins were integer valued (in particular $\pm 1$ in (\ref{tra})).
In this section, we consider the case of half-integer quantization, viz $\frac{eg}{2\pi} = \frac{1}{2}$. 
In this case, the discontinuity in phase is $\pi$ for the correlation functions (\ref{pcrc}),
and so the line of discontinuity is 
now visible. Despite this, the case is of significance. We generate operators that have spin $\frac{1}{2}$. 
They change sign under rotation by $2 \pi$. Indeed in the case of $e^2 = g^2 =\pi$ we recover  the correlation 
functions of the free fermions \cite{tsvelik,senechal}. 

The action for the free massless (Dirac) fermions in two Euclidean dimensions is: 
\bea 
S= \frac{1}{2\pi} \int d^2x \left(\bar \psi_+(x)  \partial_+ \psi_+(x) + \bar \psi_- \partial_- \psi_-(x)\right) 
\label{de} 
\eea  
where $\partial_{\pm} = \partial_1 \pm i \partial_2$.  Under rotation by an angle $\omega$, 
\bea 
\partial_{\pm} \rightarrow e^{{\mp}i\omega} \partial_{\pm}. 
\eea 
The corresponding transformations on $\psi_\pm(x)$ that leave the action (\ref{de}) invariant are: 
\bea 
\psi_{\pm}(x) \rightarrow e^{\pm i\frac{\omega}{2}} \psi_{\pm}(x), ~~~ 
\bar \psi_{\pm}(x) \rightarrow e^{\pm i\frac{\omega}{2}} \bar \psi_\pm(x). 
\label{hatp} 
\eea 
Thus $\psi_+(x)$ and $\bar \psi_+(x)$ transform the same way by half the angle. 
($\psi$ and $\bar \psi$ are independent Grassmann fields in the Euclidean functional integral). 
Thus $\psi_{\pm}, \bar \psi_\pm$ change by a sign under full rotation by $2 \pi$ and carry 
spin $\pm \frac{1}{2}$.  The propagators: 
\bea 
\langle \psi_+(x)  \bar \psi_+(y) \rangle = 2\pi  \langle x| \frac{1}{\partial_+} | y\rangle 
=- 2\pi \partial_- \bigtriangleup(x-y) = \frac{1}{(x-y)_+} =  
\frac{e^{-i\theta(\vec x - \vec y)}}{|\vec x - \vec y|}.  
\label{fpr1} 
\eea 
Here $x_{\pm} = x_1\pm ix_2$ and $\theta(\vec x)$ is the angle subtended by $\vec x$ with the $x_1$-axis. 
We also have: 
\bea 
\langle \psi_-(x) \bar \psi_-(y)\rangle =  \frac{1}{(x-y)_-} =  \frac{e^{i\theta(\vec x -\vec y)}}
{|\vec x - \vec y|}.  
\label{fpr2} 
\eea 
All other correlation functions  vanish: 
\bea 
\langle \psi_-(x) \bar \psi_+(y) \rangle =0, 
\langle \psi_+(x) \bar \psi_-(y) \rangle =0, 
\langle \psi_-(x)  \psi_-(y) \rangle =0, 
\langle \psi_+(x)  \psi_+(y) \rangle =0. 
\label{cfv} 
\eea 
The $2n$-point functions are obtained by Wick's theorem for fermions: 
\bea
\label{npcf}
&& \hspace{-0.8cm} \langle
\psi_+(x^1) \bar \psi_+(y^1) 
\psi_+(x^2) \bar \psi_+(y^2) 
\cdots 
\psi_+(x^n) \bar \psi_+(y^n) 
\rangle   =   
\frac{1}{(x^1-y^1)_+} 
\frac{1}{(x^2-y^2)_+} 
\frac{1}{(x^3-y^3)_+} \cdots 
\frac{1}{(x^n-y^n)_+}  
\nonumber   \\ &&  \hspace{-0.8cm} 
- \frac{1}{(x^1-y^2)_+} 
\frac{1}{(x^2-y^1)_+} 
\frac{1}{(x^3-y^3)_+} \cdots 
\frac{1}{(x^n-y^n)_+}  +  \cdots  
{\textrm{all possible permutations of $y^1,y^2,\cdots y^n$.}}  \nonumber\\ 
\eea 
The right hand side of (\ref{npcf}) contains $n!$ terms and each term  appears with $\pm$ sign depending upon 
the corresponding sign of permutation of $(y^1,y^2,\cdots y^n)$. Therefore, the above correlation functions 
are simply: 
\bea 
&& \hspace{-0.8cm} \langle \psi_+(x^1) \bar \psi_+(y^1) \psi_+(x^2) \bar \psi_+(y^2) \cdots 
\psi_+(x^n) \bar \psi_+(y^n) 
\rangle  
=   det\left[\frac{1}{\left(x^i-y^j\right)_+}\right]_{1 \le i,j\le n} 
\label{detcf}
\eea
This can be further written as the Cauchy determinant formula (see equation (12.195) of \cite{cft}): 
\bea 
&& \hspace{-0.8cm} \langle
\psi_+(x^1) \bar \psi_+(y^1) 
\psi_+(x^2) \bar \psi_+(y^2) 
\cdots 
\psi_+(x^n) \bar \psi_+(y^n) 
\rangle   =   
(-1)^{\frac{n(n-1)}{2}}\frac{\prod_{i <j} \left(x^i-x^j\right)_+(y^i-y^j)_+}{\prod_{i,j}\left(x^i-y^j\right)_+}  
\label{detcf1} 
\eea 
We have similar correlation functions for $\psi_-,\bar \psi_-$ with $x_+,y_+$ replaced by $x_-,y_-$. 

{\it All these correlation functions are reproduced by combinations of vertex operators of $\phi$ and the dual 
field $\tilde \phi$ ~`close together' in the special case of} 
\bea 
e^2=g^2=\pi\,.
\label{neqn}
\eea 
To show this we consider the composite operators: 
\bea 
\Psi^{\vec{\epsilon}}_{\pm}(\vec x) & \equiv & :e^{~ie \phi(\vec x+\vec\epsilon)}: 
~ :e^{~\pm ig  \tilde \phi(\vec x)}: e^{\mp ieg 
\left(\frac{\theta(\vec \epsilon)}{2\pi} + \frac{{\rm sgn} (\epsilon_2)}{4} -
\frac{1}{4}\right)}\,,    
\nonumber \\
\bar\Psi^{\vec{\epsilon}}_{\pm}(\vec x) & \equiv & :e^{~-ie \phi(\vec x+\vec\epsilon)}: 
~ :e^{~\mp ig  \tilde \phi(\vec x)}: e^{\mp ieg 
\left(\frac{\theta(\vec \epsilon)}{2\pi} + \frac{{\rm sgn} (\epsilon_2)}{4} -
\frac{1}{4}\right)}\,. 
\label{identn} 
\eea 
$ \Psi^{\vec\epsilon}_{\pm}(x)$ in (\ref{identn}), for the case (\ref{neqn}), 
correspond to a point source charge $e (=+ \sqrt{\pi})$ 
at $\vec x+\vec\epsilon$ for the field $\phi$ and $g (=\pm \sqrt{\pi})$ for the dual field $\tilde \phi$ at a nearby 
point $\vec x $.  These are the analogues of the order-disorder composites 
$\sigma(n) \mu(n^*)$ in the 2-d Ising model case 
($n$ and $n^*$ being points on the lattice and the dual lattice respectively)
which are known to behave as fermions \cite{kadceva}. 
Note that both $\Psi^{\vec\epsilon}_+ (\Psi^{\vec\epsilon}_-)$ and 
$\bar \Psi^{\vec\epsilon}_+ (\bar \Psi^{\vec\epsilon}_-)$ have the same, 
and not opposite, phase factor. This 
fact leads to the rotation property by half angle under $\theta(\vec \epsilon) \rightarrow \theta(\vec \epsilon) 
+ \omega$ just like  $\psi_+ (\psi_-)$ in (\ref{hatp}).  
Under rotation by $2\pi$, $\Psi^{\vec\epsilon}_\pm$ and 
$\bar \Psi^{\vec\epsilon}_\pm$ all change by a sign  
as $\theta(\vec \epsilon)$ changes by $2\pi$ and ${\rm sgn}(\epsilon_2)$ 
does not change.

As explained in Section \ref{sec2.3}, the field $\phi$ is non-locally related via (\ref{fp}) to the the momentum 
$\tilde\pi=\partial_2\tilde\phi$ conjugate to $\tilde\phi$. 
So $\phi(x_1,x_2)=-\int^{x_1}_{-\infty}dx_1'\,(\partial/\partial x_2) \tilde\phi(x_1',x_2)$. Thus
with $e^2=g^2=\pi$, the operators $\Psi^{\vec{\epsilon}}_{\pm}$ in (\ref{identn}) 
correspond to Mandelstam's construction of the
fermion operator in the case of the free Fermi field (as given by equations (2.8) of \cite{mandel}
for the case $\beta^2=4\pi$.) In (\ref{identn}), we give the prescription for handling this
composite fermion operator in our formalism.

From (\ref{pcrc}) and (\ref{identn}) we get: 
\bea 
\langle \Psi^{\vec\epsilon}_+(x) \bar \Psi^{\vec\epsilon^{\,\prime}}_+ (y)\rangle  = 
\frac{1}{|\vec x-\vec y|^{\frac{e^2+g^2}{2\pi}}}\,
e^{-i eg \left(\frac{\theta(\vec x-\vec y)}{\pi} + \frac{{\rm sgn}(x_2-y_2)}{2} -\frac{1}{2} \right)} \,. 
\label{crf1} 
\eea 
Note that the phase factors in (\ref{identn}) have been chosen to cancel out
the contribution of the self-correlation 
$\langle\phi(\vec x+\vec\epsilon)\tilde\phi(\vec x)\rangle$
within each composite operator (compare (\ref{pcrc})). 
Thus the correlations of $\Psi^{\vec \epsilon}$ defined as in (\ref{identn}) are
insensitive to $\vec{\epsilon}$, and we drop this superscript henceforth. 
For the special case (\ref{neqn}) we get: 
\bea 
\langle \Psi_+(x) \bar \Psi_+ (y)\rangle  =  
 ~{\rm sgn}(x_2-y_2) 
\frac{e^{-i\theta(\vec x-\vec y)}}{|\vec x-\vec y|} 
\label{wsm} 
\eea 
as $e^{i\frac{\pi}{2} {\rm sgn}(x_2-y_2)} = i {\rm sgn}(x_2-y_2)$.  
We can now reproduce the fermion propagator (\ref{fpr1}) 
with the identification: 
\bea 
\langle \psi_+(x) \bar \psi_+ (y)\rangle \equiv   ~{\rm sgn}(x_2-y_2) \langle \Psi_+(x) \bar \Psi_+ (y)\rangle. 
\label{idfp} 
\eea  
The correlator $\langle \bar \psi_+ (y)\psi_+(x)\rangle$
is identified as the negative of the right-hand side of (\ref{idfp}), as $\psi$'s are Grassmann variables.

{\it The factor  ${\rm sgn}(x_2-y_2)$ provides the crucial anti-symmetric term to relate the bosonic correlation functions 
to fermionic ones. The Klein factor required in bosonization \cite{klein} comes from this factor.} 
Consider the multi-fermion correlation function 
\bea
\langle\Psi_+(x^1) \bar \Psi_+(y^1)\Psi_+(x^2)\bar \Psi_+(y^2) \cdots \Psi_+(x^n) \bar \Psi_+(y^n)\rangle\,.                    \label{nPsi}
\eea
As $\Psi_+$, $\bar\Psi_+$ are bosonic operators, (\ref{nPsi}) is invariant under
permutations of $\{x^i\}$ and also of $\{y^i\}$. However, when we express a
correlator like $\langle\phi(x^i)\tilde\phi(x^j)\rangle$,
which involves the symmetric combination
$\theta(\vec x^i-\vec x^j)+\theta(\vec x^j-\vec x^i)$, in terms of
$\theta(\vec x^i-\vec x^j)$ alone, the resulting expression involves
the asymmetric term ${\rm sgn}(x_2^i-x_2^j)$.
In this way, (\ref{nPsi}) gets split into parts which are individually non-invariant under
permutations of $\{x^i\}$ (or of $\{y^i\}$).
 Thus, using (\ref{pcrc}), the correlator
(\ref{nPsi}) is equal to (with all self-correlations within each composite operator cancelling out)
\bea
\frac{
      \prod_{i <j} \left(x^i-x^j\right)_+
{\rm sgn}\left(x^i_2-x^j_2\right)
      \prod_{k<l}\left(y^k-y^l\right)_+
{\rm sgn}\left(y^k_2-y^l_2\right)}
     {\prod_{i,k}\left(x^i-y^k\right)_+
{\rm sgn}\left(x^i_2-y^k_2\right)}\,.
\eea
We can now reproduce (\ref{detcf1}), by identifying
\bea
&&\langle\psi_+(x^1) \bar \psi_+(y^1)\psi_+(x^2)\bar \psi_+(y^2) \cdots \psi_+(x^n) \bar \psi_+(y^n)\rangle\nonumber\\
&\equiv& (-1)^{n(n-1)/2}
\prod_{i<j}{\rm sgn}\left(x^i_2-x^j_2\right)
\prod_{k<l}{\rm sgn}\left(y^k_2-y^l_2\right)
\prod_{i,k}{\rm sgn}\left(x^i_2-y^k_2\right)\nonumber\\
&&\times\langle\Psi_+(x^1) \bar \Psi_+(y^1)\Psi_+(x^2)\bar \Psi_+(y^2) \cdots \Psi_+(x^n) \bar \Psi_+(y^n)\rangle\,.                           \label{npsi}
\eea
Let us first choose $\{x^i\}$ and $\{y^i\}$ such that
\bea
x^1_2>x_2^2>x^3_2>\cdots>x^n_2~~{\rm and}~~
y^1_2>y_2^2>y^3_2>\cdots>y^n_2\,.                        \label{choice}
\eea
In this case, all ${\rm sgn}\left(x^i_2-x^j_2\right)=+1$ for $i<j$ and
all ${\rm sgn}\left(y^k_2-y^l_2\right)=+1$ for $k<l$. Then any other ordering of
$\{x^i\}$ and $\{y^i\}$ on the left-hand side of (\ref{npsi}) will change it by
${\rm sgn}{\cal P}_x$${\rm sgn}{\cal P}_y$, where ${\rm sgn}\cal P$ is the
sign of the permutation $\cal P$ and
${\rm sgn}{\cal P}_x$ (${\rm sgn}{\cal P}_y$) is the permutation which takes
($x^1,x^2,\cdots x^n$) (respectively ($y^1,y^2,\cdots y^n$)) to the choice
(\ref{choice}).

We can also check that a correlator like
$\langle \Psi_+(x^1) \bar \Psi_+(y^1)\Psi_-(x^2)\bar \Psi_-(y^2)\rangle$
factorizes into \\
$\langle \Psi_+(x^1) \bar \Psi_+(y^1)\rangle$ and $\langle\Psi_-(x^2)\bar \Psi_-(y^2)\rangle$.
This happens because terms such as
$|\vec x^1-\vec y^2|$ involving arguments of $\Psi_+$ and $\Psi_-$ appear
in both numerator and denominator and hence cancel.
Also the phase factors such as
$\theta(\vec x^1-\vec y^2)$ cancel because the combinations
$e_i g_k$ have opposite signs.

Finally, we note that 
the two fermion number conservations, i.e.\ conservation of $\psi_+$ and of $\psi_-$,  can be traced to the two 
global symmetries (\ref{sgt}) of the bosonic fields $\phi(x)$ and $\tilde \phi(x)$
\cite{senechal}. 
As a consequence of these symmetries,
the correlations
$\langle \prod_i e^{ie_i\phi(x^i)}  \prod_k e^{ig_k \tilde \phi(x^k)}\rangle $ 
are zero  unless $\sum_i e_i = 0$ and $\sum_k g_k =0$. 
Thus these charge neutrality conditions are valid even for a finite area,
though in Section {\ref{vopdp} we obtained them by requiring infrared finiteness.  

\section{Discussion} 
\label{Discussion}

Our aim in this paper is to develop techniques for handling a field and its dual
on an equal footing. We have both fields present in a local, self-dual action. 
The dual field is interpreted in terms of line discontinuities in the configurations
of the original field (or vice versa). 
We have emphasized that the dual field explores new
configurations of the original field. In many contexts the dual field
and also composite operators made of both fields play an important role. Here we
illustrate this in the simple case of a free massless real scalar field in
two Euclidean dimensions. The action we consider has been used in string theory and is also 
related to Hamiltonians useful in
one-dimensional condensed matter physics.

This action is not manifestly invariant under rotations. The underlying
reason is the specific choice of the line discontinuity to be made. This also
gives unusual features of the correlations of the field with its dual. Even
though the correlation of either field with itself is  as expected
for scalar fields, the mutual correlations have a dependence on the orientation.
 This leads to the generation of spin from scalar fields. This is
in analogy with the Saha-Wilson contribution of the
electromagnetic field to the angular momentum of a charge-monopole system. For
correlations of vertex operators, rotation covariance is recovered at the cost
of assigning a spin to these operators. Demanding single valuedness of the
correlaton  functions, or equivalently, invisibility of the line discontinuity,
gives quantization condition for the charges of the field and the dual field. This
parallels Dirac's argument for a charge-monopole system. For  half-integer
quantization, a system of point sources for the field and the dual field close
together behaves as a fermion, in close analogy with the situation in 2-d
Ising model. 

The formalism used by us reproduces Mandelstam's
construction of the fermion operator 
and clarifies sticky points in the
bosonization programme. We have precise techniques for handling composite
operators. We throw light on how antisymmetric correlation functions of
fermions can be related to the bosonic correlation functions. 
Kinematical factors of ${\rm sgn}(x_2-y_2)$ which relate the two are
naturally obtained, giving
the Klein factors required in bosonization.

We now explain how our line discontinuity is related to soliton of zero width in the sine-Gordon
model. For this we regard the $x_1$-axis as the time direction and the $x_2$ axis as the space direction.
A point source of unit charge for the field $\phi$ at the point $\vec y$ 
produces a singular flux line, starting at $\vec y$ and
along the positive $x_1$-direction, in the partition function in
terms of the dual field $\tilde \phi$.
So there is a discontinuity 
$\Delta \tilde \phi=1$ in the configurations of $\tilde \phi$
at all times $x_1 > y_1$ at the spatial point $x_2=y_2$.
 Thus a soliton configuration of zero width is generated at time $y_1$ by the
vertex operator $e^{i\phi(y)}$.

The dual field is non-locally related to the conjugate momentum. This is the
reason that even with double the number of fields, we are not introducing new degrees
of freedom but only reinterpreting the original model.
 Even though we considered a free theory here our framework can accomodate interacting
theories as well. The crucial point is that
the dual field is related to the conjugate momentum and therefore
involves only the kinetic energy part of the action and not the potential
energy part. Thus for the theory ${\cal L}=\frac{1}{2}(\partial_i\phi)^2+ V(\phi)$ our local theory 
with both the field and its dual is 
simply: 
$${\cal L} =\frac{1}{2} (\partial_1\phi)^2 + \frac{1}{2} (\partial_1 \tilde \phi)^2 
+i \partial_1\phi(x) \partial_2 \tilde \phi(x) + V(\phi).$$
The presence of $V(\phi)$ makes important differences.  In order that the action be finite in the 
presence of the line discontinuity, $\phi$ can only take the values corresponding to the minima of $V(\phi)$.  
For $V(\phi) = \lambda (\phi^2-v^2)^2$, these are only $\pm v$ and not arbitrary as in the absence of 
$V(\phi)$. Also there is only one global invariance $\tilde \phi \rightarrow \tilde \phi + \tilde \sigma$ 
and only one conserved fermion number.

We also want to emphasize that our considerations are not restricted to two dimensions or scalar theories.
In Ref.\ \cite{plb}, we have developed local field theoretical formulation with both the field and its dual
simultaneously present for the case of Abelian gauge theory in three dimensions and also in four dimensions. 
We can handle even non-abelian gauge theories in three and four dimensions. This will be explored elsewhere. 

\section*{Acknowledgement}

I.M. thanks UGC (DRS) for support.

\vspace{1cm}

\noindent {\bf \large Appendix}

\vspace{0.4cm}

\noindent {\bf \Large On the choice of the line of discontinuity}

\noindent In Section 2, when solving the Gauss law (\ref{sgl}), we chose the line of discontinuity to be 
along the $x_1$ axis. We could have chosen an arbitray path to carry the conserved  flux.
The general solution of the Gauss law constraint is: 
\bea 
b_i(x) = \epsilon_{ij} \partial_j\tilde \phi(x) + \beta_i(x)
\label{gcs} 
\eea 
where 
\bea 
\beta_i(x) = - \int_{-\infty}^{0} d\tau \frac{dX_i(\tau)}{d\tau} \delta^2\left(x-X(\tau)\right)
\label{gfl} 
\eea 
brings the flux from $\vec X(-\infty)$ to $\vec y$ along an arbitrary path $\vec X(\tau)$ 
parametrized by a parameter $\tau$.  As $\vec y$ is the end point, $\vec X(0) = \vec y$. 
\begin{figure}[t]
\centering
\includegraphics[width=6cm,height=6cm]{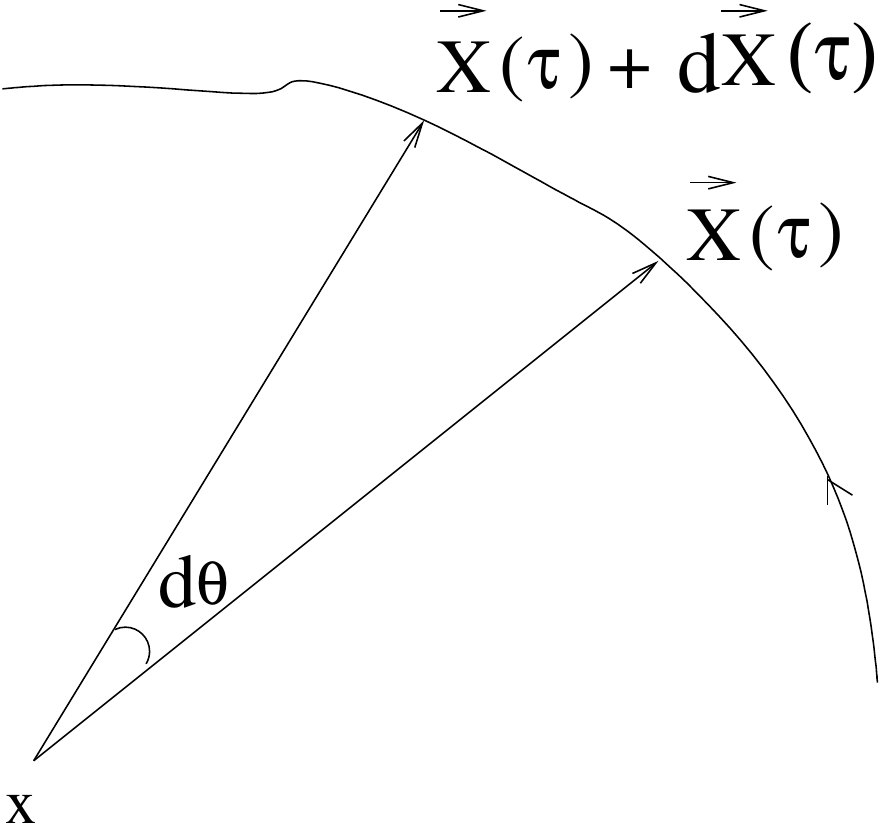}
\caption{An arbitrary line of discontinuity.} 
\label{fig2}
\end{figure}
Indeed, 
\bea 
\partial_i \beta_i(x) & = & 
- \int_{-\infty}^{0} d\tau \frac{dX_i}{d\tau} \frac{\partial}{\partial x_i} \delta^2\left(x - X(\tau)\right) 
= \int_{-\infty}^{0} d\tau \frac{dX_i}{d\tau} \frac{\partial}{\partial X_i} \delta^2\left(x - X(\tau)\right)
\nonumber \\ 
& = & \int_{-\infty}^{0} d\tau \frac{d}{d\tau} \delta^2\left(x - X(\tau)\right) = \delta^2(\vec x -\vec y) - 
\delta^2\left(\vec x - \vec X(-\infty)\right) = \delta^2\left(\vec x - \vec y\right).  
\label{deg} 
\eea 
In the last step above we have taken the other end point $\vec X(-\infty)$ of the flux line to be located outside the region 
of interest.  For different point sources located at  $\vec y^r$ ($r=1,2,\cdots N)$ we may choose  different 
discontinuity lines $\vec X^{r}(\tau)$ ending at $\vec y^{r}$. For a distributed source $\rho(x)$, 
\bea 
\beta_i(x) = - \sum_r \int_{-\infty}^{0} d\tau \frac{dX_i^r(\tau)}{d\tau} \delta^2(\vec x - \vec X^r\left(\tau)\right)   
\label{gbf} 
\eea 
with the boundary conditions $\vec X^r (\tau=0) = \vec y^{r}$.  Using (\ref{gcs}) in (\ref{dpf}) and introducing 
source term for $\tilde \phi$ we get: 
\bea 
Z[\rho,\tilde \rho] = \int D\tilde \phi \,e^{\int d^2x\left(-\frac{1}{2}\left(\epsilon_{ij}\partial_j\tilde \phi(x)+
\beta_i(x)\right)^2+i~\tilde \rho(x)\tilde \phi(x)\right)}.
\label{lpf}
\eea 
Integration over $\tilde \phi$ gives, 
\bea 
Z[\rho,\tilde \rho] = e^{-\frac{1}{2} \int d^2x \vec \beta^2(x) + \frac{1}{2}\int d^2xd^2y
\left[\partial_1\beta_2(x) -\partial_2\beta_1(x)+ i~\tilde \rho(x) \right] \bigtriangleup(x-y) 
\left[\partial_1\beta_2(y) -\partial_2\beta_1(y)+ i~\tilde \rho(y) \right]}.  
\label{123} 
\eea 
Using $\partial_i \beta_i(x)=\rho (x)$, we get
\bea
Z[\rho,\tilde \rho]=e^{\int d^2x d^2y \left(-\frac{1}{2}\rho(x) \bigtriangleup(x-y) \rho(y)  
-\frac{1}{2}\tilde \rho(x) \bigtriangleup(x-y) \tilde \rho(y) + 
{i} \tilde \rho(x) \bigtriangleup(\vec x- \vec y)\left(\partial_1\beta_2(y) -\partial_2\beta_1(y)\right)\right)}. 
\label{maeq1} 
\eea
Now
\bea 
\label{vce} 
\int d^2x d^2y \tilde \rho(x) \bigtriangleup(\vec x- \vec y)\left(\partial_1\beta_2(y) 
-\partial_2\beta_1(y)\right)  
\hspace{5cm}  \\
\hspace{4cm}  = - \frac{1}{2\pi} \int d^2x d^2y \tilde \rho(x) \int_{-\infty}^{0} d\tau \frac{dX_i}{d\tau} \epsilon_{ij}
\frac{\left( x_j  - X_j(\tau)\right)}{\left(\vec x - \vec X(\tau)\right)^2} \rho(y). \nonumber  
\eea 
We also have 
\bea 
\epsilon_{ij}\frac{\left(X_i(\tau)-x_i\right)} {\left(\vec X(\tau) - \vec x\right)^2} d \vec X_j(\tau) = d\theta 
\label{rpmn} 
\eea 
where $d\theta$ is the angle subtended by $d\vec X$ at $\vec x$,
as in Figure \ref{fig2}, with the following sign convention: 
$d\theta$ is positive (negative) according as the angle between $(\vec X - \vec x)$ and $d \vec X$  (measured in the 
anti-clockwise sense from the former to latter) is $>\pi$  ($ < \pi$).
This convention is shown in Figure \ref{fig3}. 
\begin{figure}[t]
\centering
\includegraphics[width=8cm,height=8cm]{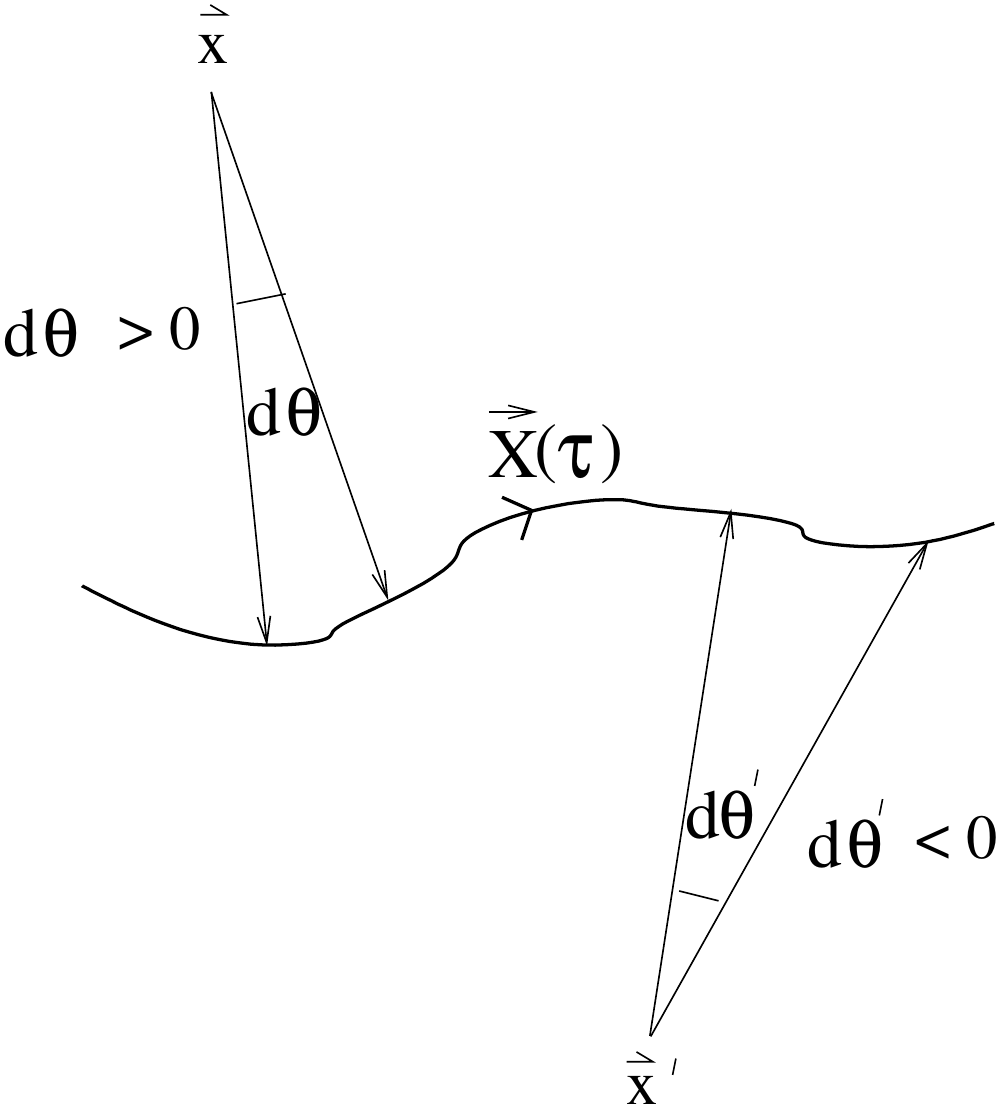}
\caption{Sign convention for $d\theta$ defined in (\ref{rpmn}) } 
\label{fig3}
\end{figure}
Thus (\ref{vce}) becomes: 
\bea 
\frac{1}{2\pi} \int d^2x d^2 y \tilde \rho(x) \theta(x-y) \rho(y) 
\eea
where 
\bea 
\theta(x-y) = \int_{-\infty}^0 d\theta({\rm P}) \label{asym}
\eea
is the net change in the angle $\theta$ along the infinite path P.  This effectively measures the angle 
from the asymptote of the path. If we choose the negative $x_1$-axis along this asympote, then (\ref{asym})
is the same as $\Theta(\vec x-\vec y)$ of Section {\ref{sec2.4}. We want to point out that here
we got only $\Theta(\vec x-\vec y)$,  and not the symmetric
combination $\frac{1}{2}(\Theta(\vec x-\vec y)+\Theta(\vec y-\vec x))$  as in (\ref{ang}), 
because we have not explicitly introduced the field $\phi$
and integrated over a symmetric quadratic form involving $\phi$ and $\tilde\phi$.
 


\end{document}